\documentclass[12pt]{article}
\usepackage{amsmath,amssymb,array}
\numberwithin{equation}{section}

\begin{document}

\begin{titlepage}
\renewcommand{\thefootnote}{\fnsymbol{footnote}}

\begin{center}
\begin{flushright}hep-th/0503089\end{flushright}
\vspace{1.5cm}

\textbf{\Large{Flat Currents of the Green-Schwarz\\[0.5cm]
Superstrings in $AdS_5\times S^1$  and $AdS_3\times S^3$
backgrounds}}\vspace{2cm}

Bin Chen$^\mathrm{a}$\footnote{Email: bchen@itp.ac.cn}\hspace{0.5cm}
Ya-Li He$^\mathrm{b}$\footnote{Email: ylhe@gms.phy.pku.edu.cn}\hspace{0.5cm}
Peng Zhang$^\mathrm{c}$\footnote{Email: pzhang@itp.ac.cn}\hspace{0.5cm}
Xing-Chang Song$^\mathrm{b}$\footnote{Email: songxc@pku.edu.cn}
\vspace{1cm}

a. \emph{Interdisciplinary Center of Theoretical Studies, Chinese Academy of Science,\\
    P.O. Box 2735, Beijing 100080, P.R. China}\\[0.3cm]
b. \emph{School of Physics, Peking University, Beijing 100871, P. R. China}\\[0.3cm]
c. \emph{Institute of Theoretical Physics, Chinese Academy of Sciences,\\
   P.O. Box 2735, Beijing 100080, P.R. China}

\end{center}\vspace{1.5cm}

\centerline{\textbf{Abstract}}\vspace{0.5cm}
We construct a one-parameter family of  flat currents in $AdS_5\times S^1$ and
$AdS_3\times S^3$ Green-Schwarz superstrings, which would
naturally lead to a hierarchy of classical conserved nonlocal
charges.   In the former case we rewrite the $AdS_5\times S^1$
string using a new $\mathbb{Z}_4$-graded base of the superalgebra
$su(2,2|2)$. In both cases the existence of the $\mathbb{Z}_4$ grading in the
superalgebras  plays a key role in the construction.
As a result, we find that the flat currents, when formally written in terms
of the $\mathcal{G}_0$-gauge invariant lowercase 1-forms, take the same form as
the one in $AdS_5\times S^5$ case.

\end{titlepage}
\setcounter{footnote}{0}


\section{Introduction}
The AdS/CFT correspondence provides a lot of deep insights into both
string theories and gauge theories \cite{M97,GKP98,W98}. This
duality realizes a fairly old idea, originated from 't Hooft
\cite{tH74}, that a large N gauge theory should be equivalent to a
string theory. A main progress in this subject in recent years is
the discovery of integrable structures on both sides. It was found
\cite{MZ02,Bei03,BeiS03} that the one-loop dilatation operator can
be interpreted as the Hamiltonian of an integrable spin chain. On
the string side, a hierarchy of infinite nonlocal charges are
constructed \cite{BPR03} in the Green-Schwarz superstring on
$AdS_5\times S^5$ spacetime \cite{MT98}, implying that the
world-sheet sigma model is integrable. Subsequently it was shown
\cite{V03} that such charges also exist in the pure spinor
formalism. The analog nonlocal charges were also found \cite{Al03}
in the ten dimensional pp-wave background and they can be identified
as the Penrose limit of the nonlocal charges in $AdS_5\times S^5$
string. 
It was shown  in \cite{DNW03} that the integrability unravelled in
the string side could be intimately connected with the Yangian
symmetry constructed in the $\mathcal{N}=4$ superconformal
Yang-Mills theory in the free-field limit (But see also \cite{Hou}).
Actually, AdS/CFT correspondence implies that the Yangian symmetry
should persist in the $\mathcal{N}=4$ superconformal Yang-Mills
theory in the planar limit even when $g^2_{YM}N$ is nonvanishing. By
using the supertwistor correspondence \cite{W03} of the
$\mathcal{N}=4$ SYM, the author of \cite{Wolf04} studied the
supertwistor construction of hidden symmetry algebras of the
self-dual SYM equations.  Recently, in the pure spinor formalism of
superstrings on $AdS_5\times S^5$, it was proved in \cite{Berk01}
that the nonlocal charges in the string theory are BRST-invariant
and physical,  and in \cite{Berk02} it was shown that there exists
an infinite set of nonlocal BRST-invariant currents at the quantum
level. Some recent developments on nonlocal charges could be found
in \cite{BKSZ05, Tsey2005}.

It is important to study the AdS/CFT correspondence in the cases
with less supersymmetry, in order to better understand the
correspondence and be close to the real world. There are various
ways to partially break the supersymmetry. By calculating the
1-loop $\beta$ function Polyakov \cite{Poly99} proposed that the
noncritical $AdS_p\times S^{\,q}$ string theories should be dual
to the gauge theories with less or no supersymmetry. Last year he
briefly constructed several sigma models in \cite{Poly04} and
argued that they are conformal invariant and also completely
integrable \footnote {In \cite{Poly04} the author asserted that
the sigma models described in that paper are completely integrable
with explicit construction of the flat currents in a
$OSp\,(2|4)/(SO(3,1)\times SO(2))$ model.}, just as the critical
case. Soon after, Klebanov and Maldacena \cite{KM04} found the
$AdS_5\times S^1$ solution  in the low energy supergravity
effective action of six dimensional noncritical string theory with
$N$ units of RR flux and in the presence of $N_f$ space-time
filling $D5$-branes. This solution has the right structure to be
dual to $\mathcal{N}=1$ supersymmetric $SU(N)$ gauge theories with
$N_f$ flavors, in agreement with the proposal in \cite{Poly04}.
 Several other
$AdS_p\times S^{\,q}$ solutions were also found \cite{3Iranian} in
the context of the six dimensional noncritical superstring theory.

In this paper, as the first step to understand the integrable
structure in Green-Schwarz superstring on $AdS_5\times S^1$ \cite{Poly04},
we try to construct the flat currents of this
model, which would naturally lead to infinite nonlocal charges. It
has been shown that the coset structure and the $\mathbb{Z}_4$
grading of $psu(2,2|4)$, which is the symmetry algebra of the
$AdS_5\times S^5$ string, are essential to construct the nonlocal
charges\cite{BPR03}. The Green-Schwarz superstring on $AdS_5\times
S^1$ is a sigma model with the target space
$SU(2,2|2)/(SO(4,1)\times SO(3))$. To find the flat currents we need
first determine whether the superalgebra $su(2,2|2)$ has the similar
grading structure. Unlike $su(2,2|4)\simeq psu(2,2|4)\oplus u(1)$
can be decomposed into a direct sum of two algebras, the algebra
$su(2,2|2)$ is already simple by itself. The $u(1)$ part has the
non-zero (anti)commutators with other generators of $su(2,2|2)$. In
spite of this difference we find that the algebra $su(2,2|2)$ also
has a $\mathbb{Z}_4$ grading structure by explicitly constructing
the graded generators, and that the denominator of the supercoset
$SU(2,2|2)/(SO(4,1)\times SO(3))$ corresponds to the part of grade
zero. In terms of these new graded generators, we rewrite the
$\kappa$-symmetric action of the $AdS_5\times S^1$ string and derive
the equations of motions. The construction of the flat currents
follows \cite{BPR03} in a straightforward way. In terms of the
$\mathcal{G}_0$-gauge invariant lowercase forms, the Maurer-Cartan
equations and the equations of motion take the same form as the ones
in the $AdS_5\times S^5$ string. As a result, the flat currents we
construct, when written in terms of the gauge invariant 1-forms,
have the same form as those of the $AdS_5\times S^5$ case.

Another supercoset model with $\mathbb{Z}_4$ grading is the
Green-Schwarz superstring in $AdS_3\times S^3$ background, which
was constructed in \cite{RR98}. We construct the flat currents and
find it also has the same form as the ones  in the $AdS_5\times
S^5$ string.

This paper is organized as follows. We first study the
$AdS_5\times S^1$ string. In section 2 we construct the
$\mathbb{Z}_4$-graded generators of the superalgebra $su(2,2|2)$
explicitly, then we define the Maurer-Cartan 1-forms with respect
to these generators and write down the Maurer-Cartan equations
that these forms must satisfy. In section 3 we rewrite the action
of $AdS_5\times S^1$ string using the graded generators and then
get the equations of motion. Then we explicitly construct
a one-parameter family of  the flat currents in $AdS_5\times S^1$
superstring. In section 4 we construct the flat currents in
$AdS_3\times S^3$ superstring. In section 5 we review our results
and discuss some future directions in this subject. In appendix A
we list our conventions in this paper, which are mainly used in
the construction of the $\mathbb{Z}_4$-graded generators of
$su(2,2|2)$. In appendix B we check the $\kappa$-symmetry of the
action of the $AdS_5\times S^1$ superstring. In appendix C we
write down the closed expressions of the Maurer-Cartan 1-forms of
the coset space $SU(2,2|2)/(SO(4,1)\times SO(3))$.


\section{$su(2,2\,|2)$ superalgebra}
Our starting point is the extended superconformal algebra $su(2,2|2)$,
which is the symmetry algebra of the $AdS_5\times S^1$ superstring.
This is a simple superalgebra.
Its bosonic part is $su(2,2)\oplus su(2)\oplus\,u(1)$. We denote
$M_{\bar{a}\bar{b}},T_{a'},R$ as the generators of
$su(2,2),su(2),u(1)$ respectively. Notice the isomorphisms
$su(2,2)\simeq so(2,4)$ and $su(2)\simeq so(3)$, then
\begin{eqnarray}
&&[M_{\bar{a}\bar{b}},M_{\bar{c}\bar{d}}]
=\eta_{\bar{a}\bar{d}}M_{\bar{b}\bar{c}}+\eta_{\bar{b}\bar{c}}M_{\bar{a}\bar{d}}
-\eta_{\bar{a}\bar{c}}M_{\bar{b}\bar{d}}-\eta_{\bar{b}\bar{d}}M_{\bar{a}\bar{c}}\,,\nonumber\\[0cm]
&&[T_{a'},T_{b'}]=\varepsilon_{a'b'c'}T_{c'}\,.\label{BB}
\end{eqnarray}
Here
$(\eta_{\bar{a}\bar{b}})=\mathrm{diag}(-++++\,\,-),\,(\eta_{a'b'})=\mathrm{diag}(+++)$
with $\bar{a},\bar{b}=0,1,2,3,4,5$ and $a',b'=1,2,3$.

The fermionic part of $su(2,2|2)$ lies in a
$(\mathbf{4,2,1})\oplus(\mathbf{4^*,2^*,1^*})$ representation of the bosonic part
under the adjoint action. If we denote $F_{\alpha\alpha'}$ and $\bar{F}^{\alpha\alpha'}$
as the fermionic generators with $\alpha,\beta=1,2,3,4$ and $\alpha',\beta'=1,2$,
then
\begin{eqnarray}
[F^{\alpha\alpha'},M_{\bar{a}\bar{b}}]
  =\frac{1}{2}(\Gamma_{\bar{a}\bar{b}})^\alpha_\beta F^{\beta\alpha'}\,, \quad&&
  [\bar{F}_{\alpha\alpha'},M_{\bar{a}\bar{b}}]
  =-\frac{1}{2}\bar{F}_{\beta\alpha'}(\Gamma_{\bar{a}\bar{b}})^\beta_\alpha\,,\nonumber\\[0cm]
[F^{\alpha\alpha'},T_{a'}]
  =\frac{1}{2}(\tau_{a'})^{\alpha'}_{\beta'} F^{\alpha\beta'}\,, \quad&&
  [\bar{F}_{\alpha\alpha'},T_{a'}]
  =-\frac{1}{2}\bar{F}_{\alpha\beta'}(\tau_{a'})^{\beta'}_{\alpha'}\,,\nonumber\\[0cm]
[F^{\alpha\alpha'},R\,]=\frac{\mathrm{i}}{2}F^{\alpha\alpha'}\,, \quad&&
  [\bar{F}_{\alpha\alpha'},R\,]=-\frac{\mathrm{i}}{2}\bar{F}_{\alpha\alpha'}\,.\label{FB}
\end{eqnarray}
Here $\Gamma_{\bar{a}\bar{b}}$ are $4\times4$ matrices, their
explicit form can be found in the appendix A, and
$\tau_{a'}=-\mathrm{i}\sigma_{a'}$ with $\sigma_{a'}(a'=1,2,3)$ being the three
Pauli matrices. In the above formulae we choose the convention
that the upper indices of $\Gamma$'s and $\tau$'s are the row
indices, and that the lower ones are the column indices.

At last the anticommutator of two fermionic generators are
\begin{eqnarray}
\{F^{\alpha\alpha'},\bar{F}_{\alpha\alpha'}\}=
-\frac{\mathrm{i}}{2}(\Gamma^{\bar{a}\bar{b}})^\alpha_\beta\, \delta^{\alpha'}_{\beta'}M_{\bar{a}\bar{b}}
-\delta^\alpha_\beta \delta^{\alpha'}_{\beta'}R
+2\mathrm{i}\delta^\alpha_\beta (\tau^{a'})^{\alpha'}_{\beta'}T_{a'}\,.\label{FF}
\end{eqnarray}
The structure of the right hand side of the above anticommutator is the consequence
of the covariance, and the coefficients in front of each terms are fixed by the
Jacobi identities. The commutation relations (\ref{BB}) (\ref{FB}) and (\ref{FF})
entirely define the superalgebra $su(2,2|2)$ in the standard way.
The author of \cite{Poly04} used just these generators to construct the action
of $AdS_5\times S^1$ superstring. The grading structure is implicit.


\subsection{Construction of the graded generators}

Now we have the commutation relations of $su(2,2|2)$, but we
cannot see clearly whether there is a $\mathbb{Z}_4$ grading in
terms of the generators $(M_{\bar{a}\bar{b}},T_{a'},R,F,\bar{F})$.
The grading structure of the symmetry algebra is an important ingredient
in the study of nonlocal charges of the $AdS_5\times S^5$ string.
In this subsection we demonstrate
the $\mathbb{Z}_4$-grading of the superalgebra $su(2,2|2)$ by
constructing the graded generators explicitly.

The fermionic generators $F^{\alpha\alpha'}$ and
$\bar{F}_{\alpha\alpha'}$ are 4-component spinors with respect to
the index $\alpha$. We split them into 2-component spinors as
$F^{\alpha\alpha'}=(U^{A\alpha'},{V_{\dot{A}}}^{\alpha'})$ and
$\bar{F}_{\alpha\alpha'}=(V_{A\alpha'},{U^{\dot{A}}}_{\alpha'})$,
where
${V_{\dot{A}}}^{\alpha'}=C^{\,\alpha'\beta'}(V_{A\beta'})^\dag$
and ${U^{\dot{A}}}_{\alpha'}=C_{\alpha'\beta'}(U^{A\beta'})^\dag$.
Next we define
$P_a=M_{a5},\, J_{ab}=M_{ab}\,(a,b=0,1,2,3,4)$ and
$Q^I_{\alpha\alpha'}=(\,Q^I_{A\alpha'}\,,{Q^{I\dot{A}}}_{\alpha'})\,\,\,I=1,2$ by
\begin{eqnarray}
Q^1_{A\alpha'}&=&\varepsilon_{AB}C_{\alpha'\beta'}U^{B\beta'}-V_{A\alpha'}\quad \nonumber\\[0.2cm]
{Q^{1\dot{A}}}_{\alpha'}
  &=&-\varepsilon^{\dot{A}\dot{B}}C_{\alpha'\beta'}{V_{\dot{B}}}^{\beta'}-{U^{\dot{A}}}_{\alpha'}\quad \nonumber\\[0.2cm]
Q^2_{A\alpha'}&=&\mathrm{i}\,(\varepsilon_{AB}C_{\alpha'\beta'}U^{B\beta'}+V_{A\alpha'})\quad\nonumber\\[0.2cm]
{Q^{2\dot{A}}}_{\alpha'}
  &=&\mathrm{i}\,(-\varepsilon^{\dot{A}\dot{B}}C_{\alpha'\beta'}{V_{\dot{B}}}^{\beta'}+{U^{\dot{A}}}_{\alpha'}).
\end{eqnarray}
In terms of the new generators
$(P_a,J_{ab},R,T_{a'},Q^I_{\alpha\alpha'})$, the commutation
relations of the superalgebra change to :
\begin{eqnarray}
&&[P_a,P_b]=J_{ab},\quad
  [P_a,J_{bc}]=\eta_{ab}P_c-\eta_{ac}P_b,\quad
  [T_{a'},T_{b'}]=\varepsilon_{a'b'c'}T_{c'}\,\,,\nonumber\\[0.2cm]
&&[J_{ab},J_{cd}]=\eta_{ad}J_{bc}+\eta_{bc}J_{ad}-\eta_{ac}J_{bd}-\eta_{bd}J_{ac}\,\,,\nonumber\\[0.1cm]
&&[Q^I_{\alpha\alpha'},P_a]=-\frac{\mathrm{i}}{2}\varepsilon^{IJ}Q^J_{\beta\alpha'}\left(\gamma_a\right)^\beta_\alpha,\quad
  [Q^I_{\alpha\alpha'},R\,]=\frac{1}{2}\varepsilon^{IJ}Q^J_{\alpha\alpha'},\nonumber\\
&&[Q^I_{\alpha\alpha'},J_{ab}]=-\frac{1}{2}Q^I_{\beta\alpha'}\left(\gamma_{ab}\right)^\beta_\alpha,\quad
  [Q^I_{\alpha\alpha'},T_{a'}]=-\frac{1}{2}Q^I_{\alpha\beta'}\left(\tau_{a'}\right)^{\beta'}_{\alpha'},\nonumber\\[0.1cm]
&&\{Q^I_{\alpha\alpha'},Q^J_{\beta\beta'}\}=\delta^{IJ}\left[-2\mathrm{i}\left(C\gamma^a\right)_{\alpha\beta}C_{\alpha'\beta'}P_a
  +2C_{\alpha\beta}C_{\alpha'\beta'}R\,\right]\nonumber\\[0.1cm]
&&\quad\quad\quad\quad\quad\quad\quad
  +\,\varepsilon^{IJ}\left[\,(C\gamma^{ab})_{\alpha\beta}C_{\alpha'\beta'}J_{ab}
  -4C_{\alpha\beta}(C\,'\tau^{a'})_{\alpha'\beta'}T_{a'}\right].\label{QQ}
\end{eqnarray}
Here $a,b=0,1,2,3,4$ which can be lowered and raised by the
5-dimensional metric
$(\eta_{ab})=(\eta^{ab})=\mathrm{diag}(-++++)$, and $a',b'=1,2,3$
which can be lowered and raised by the 3-dimensional metric
$(\,\eta_{a'b'})=(\,\eta^{a'b'})=\mathrm{diag}(+++)$. The indices
$I,J=1,2$ and the Levi-Civita symbol is defined by
$\varepsilon^{12}=-\varepsilon^{21}=1$. The gamma-matrices are
\begin{eqnarray}
\gamma^a=\begin{cases} \,\gamma^\mu,& a=\mu,\\ \,\bar{\gamma}\,,&
a=4, \end{cases} \quad\quad
\gamma^{ab}=\frac{1}{2}[\gamma^a,\gamma^b],\quad\quad
\tau^{a'}=-\mathrm{i}\,\sigma^{a'}.
\end{eqnarray}
Their explicit expressions can be found in Appendix A.
The matrices $\gamma^a$ and $\tau^{a'}$ are the generators of 5-dimensional and 3-dimensional
Clifford algebra, $C$ and $C\,'$ are their charge conjugation matrices\footnote
{To simplify the the notation, the prime on $C\,'$ is explicitly written only when it does not
carry the indices. So $C_{\alpha'\beta'}$ are the elements of the matrix $C\,'$.}
respectively, which satisfy
\begin{eqnarray}
C\gamma^a=(\gamma^a)^TC,\quad C\,'\tau^{a'}=-(\tau^{a'})^TC\,'.
\end{eqnarray}
Their explicit form can be seen in (\ref{C}) and (\ref{C'}).

Now we can see the $\mathbb{Z}_4$-grading of $su(2,2|2)$ explicitly from (\ref{QQ}). Therefore
$su(2,2|2)=\mathcal{G}_0\oplus\mathcal{G}_1\oplus\mathcal{G}_2\oplus\mathcal{G}_3$ with
\begin{eqnarray}
&&\mathcal{G}_0=\{J_{ab},T_{a'}\},\nonumber\\
&&\mathcal{G}_1=\{Q^1_{\alpha\alpha'}\},\nonumber\\
&&\mathcal{G}_2=\{P_{a},R\,\},\nonumber\\
&&\mathcal{G}_3=\{Q^2_{\alpha\alpha'}\},
\end{eqnarray}
where $\mathcal{G}_i$ denotes the component of grade $i$. The
target space of $AdS_5\times S^1$ superstring is the supercoset
space $SU(2,2|2)/\,(SO(4,1)\times SO(3))$. As in the $AdS_5\times
S^5$ case, the algebra of the isotropy group, which would be
gauged away, is just the grade zero part $\mathcal{G}_0$ of the whole
algebra. This similarity determines that the $AdS_5\times S^1$
string has many common properties with the intensely studied
$AdS_5\times S^5$ case. For instance it enables us to use the
methods of \cite{BPR03} to construct the flat currents which would
lead to the classical conserved nonlocal charges.


\subsection{Maurer-Cartan 1-forms}
It is convenient to use the Maurer-Cartan 1-forms to construct the
worldsheet action. Here our target space is the coset space
$SU(2,2|2)/\,(SO(4,1)\times SO(3))$. The Maurer-Cartan 1-form
$G^{-1}dG$ of the supergroup $SU(2,2|2)$ is pulled back to the
coset space by the local section $G=G(x,\theta)$, where the
coordinates $(x,\theta)$ parametrize the coset space and
$G(x,\theta)$ is an element of the supergroup $SU(2,2|2)$. In the
following we shall call the pull-back of the Maurer-Cartan 1-form
of the supergroup the Maurer-Cartan 1-form of the coset space.

The left invariant Maurer-Cartan 1-form $\mathbf{L}\equiv G(x,\theta)^{-1}dG(x,\theta)$
of the coset space takes its value in the superalgebra
$su(2,2|2)$, so we can expand it using the generators of this algebra
\begin{eqnarray}
\mathbf{L}\equiv G(x,\theta)^{-1}dG(x,\theta)
 \equiv P_a L^a+R L^R+\frac{1}{2}J_{ab}L^{ab}+T_{a'}L^{a'}
        +Q^I_{\alpha\alpha'}L^{I\alpha\alpha'}\,,
\end{eqnarray}
where $L^a$ and $L^R$ are the 5-beins and 1-beins of $AdS_5$ and
$S^1$ respectively. $L^{I\alpha\alpha'}$ are the two ($I=1,2$)
spinorial  8-beins which are two Majorana spinors of
$SO(4,1)\times SO(3)$. $L^{ab}$ and $L^{a'}$ are the $SO(4,1)$ and
$SO(3)$ connections respectively. The components of the Lie
algebra valued 1-form $\mathbf{L}$ are super 1-forms. They are both
differential 1-forms and the functions of Grassmann variables
$\theta$. Among these components $L^a,L^R,L^{ab}$ and $L^{a'}$ are
Grassmann even quantities, while $L^{I\alpha\alpha'}$ are
Grassmann odd ones. So we have $L^a \wedge
L^{I\alpha\alpha'}=-\,L^{I\alpha\alpha'}\wedge L^a$, but
$L^{I\alpha\alpha'}\wedge
L^{J\beta\beta'}=+\,L^{J\beta\beta'}\wedge L^{I\alpha\alpha'}$.

The Maurer-Cartan 1-form satisfies the zero-curvature equation
$d\,\mathbf{L}=-\frac{1}{2}\,[\,\mathbf{L},\mathbf{L}]$. Then we get the following Maurer-Cartan
equations
\begin{eqnarray}
&& dL^a=-L^b\wedge L^{ba}-\mathrm{i}\bar{L}^I\gamma^a\wedge L^I,\\[0.1cm]
&& dL^R=\,\bar{L}^I\wedge L^I,\\
&& dL^I=\frac{\mathrm{i}}{2}\varepsilon^{IJ}(L^a\gamma^a+\mathrm{i} L^R)\wedge L^J
      -\frac{1}{4}(L^{ab}\gamma^{ab}+2L^{a'}\tau^{a'})\wedge L^I.\label{dL^I}
\end{eqnarray}
Here we define the Majorana conjugation
$\bar{L}^I_{\beta\beta'}=L^{I\alpha\alpha'}C_{\alpha\beta}C_{\alpha'\beta'}$.
Then we have
\begin{eqnarray}
d\bar{L}^I=-\frac{\mathrm{i}}{2}\varepsilon^{IJ}\bar{L}^J\wedge(L^a\gamma^a+\mathrm{i} L^R)
      -\frac{1}{4}\bar{L}^I\wedge(L^{ab}\gamma^{ab}+2L^{a'}\tau^{a'}).
\end{eqnarray}


\section{Flat Currents of the GS superstring on $AdS_5\times S^1$}

The action of $AdS_5\times S^1$ superstring is
\begin{eqnarray}
&& S=S_0+S_1\,,\\[0.1cm]
&& S_0=-\frac{1}{2}\int d^2\sigma \sqrt{-g}\,g^{ij}(L^a_i L^a_j+L^R_i L^R_j)\,,\\
&& S_1=k\int \bar{L}^1\wedge L^2\,.
\end{eqnarray}
Here we choose the Wess-Zumino term in a quadratic form. This kind
of action has been discussed in \cite{Poly04,Hatsuda}. The
coefficient $k$ will be determined by requiring $\kappa$-symmetry
of the whole action $S$ (see appendix B). It is fixed to be $-2$.

To check the $\kappa$-symmetry and also to derive the equations of
motion, we need the variations of the Maurer-Cartan forms. Suppose
that we make the variation $G\rightarrow G\,'\simeq G(1+\omega)$
by right multiplication, where $\omega\equiv P_a
\omega^a+R\,\omega^R+\frac{1}{2}J_{ab}\,\omega^{ab}+T_{a'}\omega^{a'}
        +Q^I_{\alpha\alpha'}\omega^{I\alpha\alpha'}$ is an infinitesimal element of the algebra
(not a differential form). By definition $\mathbf{L}=G^{-1}dG$ we know
$\delta \mathbf{L}=d\omega+[\mathbf{L},\omega]$. So the variations of the
Maurer-Cartan 1-forms are
\begin{eqnarray}
&& \delta L^a=d\omega^a+L^{ab}\omega^b+L^b\omega^{ba}+2\mathrm{i} \bar{L}^I\gamma^a\omega^I\,,\label{vL^a}\\[0.2cm]
&& \delta L^R=d\omega^R-2\bar{L}^I\omega^I,\label{vL^R}\\[0.1cm]
&& \delta L^I=d\omega^I+\frac{\mathrm{i}}{2}\varepsilon^{IJ}(\,\omega^a\gamma^a+\mathrm{i}\,\omega^R)L^J
                       -\frac{\mathrm{i}}{2}\varepsilon^{IJ}(L^a\gamma^a+\mathrm{i} L^R)\omega^J \nonumber\\
&& \qquad\qquad        -\frac{1}{4}(\,\omega^{ab}\gamma^{ab}+2\omega^{a'}\tau^{a'})L^I
                       +\frac{1}{4}(L^{ab}\gamma^{ab}+2L^{a'}\tau^{a'})\omega^I.\label{vL^I}
\end{eqnarray}
From above equations we get the variation of the Wess-Zumino term $(s^{IJ}=\mathrm{diag}(1,-1)\,)$
\begin{eqnarray}
&&\delta\,(\bar{L}^1\wedge L^2)=d\,(\,\bar{\omega}^1L^2-\bar{L}^1\omega^2)
 -\frac{\mathrm{i}}{2}s^{IJ}\bar{L}^I(\,\omega^a\gamma^a+\mathrm{i}\,\omega^R)\wedge L^J\nonumber\\
&&\qquad\qquad\qquad +\,\mathrm{i} s^{IJ}\bar{\omega}^I(L^a\gamma^a+\mathrm{i} L^R)\wedge L^J\,.\label{vWZ}
\end{eqnarray}
Here we have used the variations of $\bar{L}^I$
\begin{eqnarray}
&& \delta\bar{L}^I=d\bar{\omega}^I+\frac{\mathrm{i}}{2}\varepsilon^{IJ}\bar{L}^J(\,\omega^a\gamma^a+\mathrm{i}\,\omega^R)
                       -\frac{\mathrm{i}}{2}\varepsilon^{IJ}\bar{\omega}^J(L^a\gamma^a+\mathrm{i} L^R) \nonumber\\
&& \qquad\qquad        +\frac{1}{4}\bar{L}^I(\,\omega^{ab}\gamma^{ab}+2\omega^{a'}\tau^{a'})
                       -\frac{1}{4}\bar{\omega}^I(L^{ab}\gamma^{ab}+2L^{a'}\tau^{a'}).
\end{eqnarray}

By use of equations (\ref{vL^a}) to (\ref{vWZ}) and fixing $k=-2$
we get the equations of motion in terms of the Maurer-Cartan 1-forms as following
\begin{eqnarray}
&&\sqrt{-g}\,g^{ij}(\nabla_i L^a_j+L^{ab}_i L^b_j)+\mathrm{i}\,\varepsilon^{ij}s^{IJ}\bar{L}^I_i\gamma^a L^J_j=0\,,\\[0.1cm]
&&\sqrt{-g}\,g^{ij}\nabla_i L^R_j-\varepsilon^{ij}s^{IJ}\bar{L}^I_i L^J_j=0\,,\\[0.1cm]
&&(\sqrt{-g}\,g^{ij}\delta^{IJ}-\varepsilon^{ij}s^{IJ})\,(L^a_i\gamma^a+\mathrm{i}L^R_i)L^J_j=0\,.
\end{eqnarray}
Here $\nabla_i$ is the covariant derivative with respect to the world-sheet metric $g_{ij}$.
As usual there is another constraint that should be supplemented to the above three equations
\begin{eqnarray}
L^a_i L^a_j + L^R_i L^R_j =\frac{1}{2}\,g_{ij}\,g^{kl}(L^a_k L^a_l + L^R_k L^R_l).
\end{eqnarray}
This equation is the consequence of the vanishing of $\,\delta S/\delta g^{ij}$.


\subsection{Flat currents}

In this subsection we follow the method of \cite{BPR03} to find
the flat currents which would lead to the classical nonlocal
charges of the $AdS_5\times S^1$ superstring. We decompose the
left invariant Maurer-Cartan form as
\begin{eqnarray}
-\mathbf{L}\equiv-G^{-1}dG\,=\,\mathbf{H}+\mathbf{P}+\mathbf{Q}^1+\mathbf{Q}^2
\end{eqnarray}
with
\begin{eqnarray}
&&\mathbf{H}=-\frac{1}{2}J_{ab}L^{ab}-T_{a'}L^{a'},\quad
  \mathbf{P}=-P_a L^a - R L^R,\nonumber\\
&&\mathbf{Q}^I=-\,Q^I_{\alpha\alpha'}L^{I\alpha\alpha'}\quad(\mathrm{no\,\,summation\,\,over}\,\,I)\,,\nonumber\\[0.1cm]
&&\mathbf{Q}\equiv\mathbf{Q}^1+\mathbf{Q}^2\,,\quad
  \mathbf{Q}'\equiv\mathbf{Q}^1-\mathbf{Q}^2.
\end{eqnarray}
To find the flat currents it is convenient to transform above
Lie superalgebra valued 1-forms denoted by capital letters to the ones
denoted by lowercase letters by conjugation
$\mathbf{x}=G\,\mathbf{X}\,G^{-1}$. Notice that $\mathbf{H}$
transforms as a connection under the $\mathcal{G}_0$-gauge
transformations, while $\mathbf{P},\mathbf{Q},\,\mathbf{Q}'$
transform in the adjoint representation of $\mathcal{G}_0$. So the
lowercase forms $\mathbf{p},\mathbf{q},\mathbf{q}'$ are
$\mathcal{G}_0$-gauge invariant. Although the capital forms
correspond to the decomposition of the Maurer-Cartan forms under
the grading of the superalgebra, the lowercase forms, however, do
not reflect the grading.

The Maurer-Cartan equations in terms of this lowercase quantities are
\begin{eqnarray}
d\mathbf{h}\,&=&-\mathbf{h}\wedge\mathbf{h}+\mathbf{p}\wedge\mathbf{p}
   -(\mathbf{h}\wedge\mathbf{p}+\mathbf{p}\wedge\mathbf{h})-(\mathbf{h}\wedge\mathbf{q}+\mathbf{q}\wedge\mathbf{h})
   +\frac{1}{2}(\mathbf{q}\wedge\mathbf{q}-\mathbf{q}'\wedge\mathbf{q}')\,,\nonumber\\
d\mathbf{p}\,&=&-2\mathbf{p}\wedge\mathbf{p}-(\mathbf{p}\wedge\mathbf{q}+\mathbf{q}\wedge\mathbf{p})
   +\frac{1}{2}(\mathbf{q}\wedge\mathbf{q}+\mathbf{q}'\wedge\mathbf{q}')\,,\nonumber\\[0.2cm]
d\mathbf{q}\,&=&-2\mathbf{q}\wedge\mathbf{q}\,,\nonumber\\[0.2cm]
d\mathbf{q}'&=&-2(\mathbf{p}\wedge\mathbf{q}'+\mathbf{q}'\wedge\mathbf{p})
   -(\mathbf{q}\wedge\mathbf{q}'+\mathbf{q}'\wedge\mathbf{q})\,.\label{MCL}
\end{eqnarray}
And the equations of motions could be rewritten as
\begin{eqnarray}
d\ast\mathbf{p}&=&(\,\mathbf{p}\wedge\ast\,\mathbf{q}+\ast\,\mathbf{q}\wedge\mathbf{p})
  +\frac{1}{2}(\mathbf{q}\wedge\mathbf{q}'+\mathbf{q}'\wedge\mathbf{q})\,,\nonumber\\[0.05cm]
0&=&\mathbf{p}\wedge(\ast\,\mathbf{q}-\mathbf{q}')+(\ast\,\mathbf{q}-\mathbf{q}')\wedge\mathbf{p}\,,\nonumber\\[0.1cm]
0&=&\mathbf{p}\wedge(\mathbf{q}-\ast\,\mathbf{q}')+(\mathbf{q}-\ast\,\mathbf{q}')\wedge\mathbf{p}\,.\label{EQL}
\end{eqnarray}
The Maurer-Cartan equations (\ref{MCL}) and the equations of
motions (\ref{EQL})  have the same form as the corresponding ones in
$AdS_5\times S^5$ case \cite{BPR03}. So there is also a family of
flat currents
$\mathbf{a}=\alpha\,\mathbf{p}+\beta\ast\mathbf{p}+\gamma\,\mathbf{q}+\delta\,\mathbf{q}'$
in $AdS_5\times S^1$ superstring parameterized by
\begin{eqnarray}
\alpha&=&-2\sinh^2\lambda\,,\nonumber\\
\beta&=&\mp2\sinh\lambda\,\cosh\lambda\,,\nonumber\\
\gamma&=&1\pm\cosh\lambda\,,\nonumber\\
\delta&=&\sinh\lambda\,,
\end{eqnarray}
by requiring the zero-curvature equation
$d\mathbf{a}+\mathbf{a}\wedge\mathbf{a}=0$. This one-parameter
family of flat currents would naturally lead to a hierarchy of
classical conserved nonlocal charges by the standard method
\cite{BPR03,2Jap,Das04}. This fact is a characteristic property
that the $2d$ sigma model of $AdS_5\times S^1$ is completely
integrable at least at classical level.

In \cite{Das04}, it has been argued that there exists another class of
flat currents in the $AdS_5\times S^5$ case, which respect the $\mathbb{Z}_4$ symmetry
explicitly. It was later realized \cite{Young} that the two currents are
equivalent.


\section{Flat currents of $AdS_3\times S^3$ Superstring}
The Green-Schwarz superstring in $AdS_3\times S^3$ space is
constructed in \cite{RR98}. In this section we shall use the
notation of that paper. The action is defined as a two-dimensional
sigma model with the target space $SU(1,1|2)^2/(SO(1,2)\times
SO(3))$. We define the Maurer-Cartan 1-forms of $SU(1,1|2)^2$
\begin{eqnarray}
\mathbf{L}\equiv G^{-1}d\,G
 \equiv L^{\hat{a}}P_{\hat{a}}+\frac{1}{2}\,L^{\hat{a}\hat{b}}J_{\hat{a}\hat{b}}
 +\frac{1}{2}\,\bar{L}^IQ^I+\frac{1}{2}\,\bar{Q}^IL^{I}\,.
\end{eqnarray}
A main difference between the $AdS_3\times S^3$ string and the
former $AdS_5\times S^1$ string is $\bar{Q}$ and $Q$ are
independent generators in the former case, while in the latter
case $\bar{Q}=Q(\,C\otimes C\,')$. The $SU(1,1|2)^2$ algebra has a
$\mathbb{Z}_4$-grading structure with
\begin{eqnarray}
\mathcal{G}_0&=&\{J_{ab},\,J_{a'b'}\}\,,\nonumber\\
\mathcal{G}_1&=&\{Q^1,\,\bar{Q}^1\}\,,\nonumber\\
\mathcal{G}_2&=&\{P_a,\,P_{a'}\}\,,\nonumber\\
\mathcal{G}_3&=&\{Q^2,\,\bar{Q}^2\}\,.
\end{eqnarray}
The grade zero part corresponds to the denominator
$SO(1,2)\times SO(3)$ of the supercoset. We decompose the
Maurer-Cartan 1-forms with respect to the grading
\begin{eqnarray}
&&\mathbf{H}=-\frac{1}{2}L^{\hat{a}\hat{b}}J_{\hat{a}\hat{b}}\,,\quad
  \mathbf{P}=-L^{\hat{a}} P_{\hat{a}}\,,\quad
  \mathbf{Q}^I=-\frac{1}{2}\bar{L}^IQ^I-\frac{1}{2}\bar{Q}^IL^I.\nonumber\\
&&\mathbf{Q}=\mathbf{Q}^1+\mathbf{Q}^2\,,\quad
  \mathbf{Q}'=\mathbf{Q}^1-\mathbf{Q}^2\,.
\end{eqnarray}
and define the corresponding lowercase forms
$\mathbf{x}=G\,\mathbf{X}\,G^{-1}$ which will be used to construct
the flat currents. They satisfy the following Maurer-Cartan
equations
\begin{eqnarray}
d\mathbf{h}\,&=&-\mathbf{h}\wedge\mathbf{h}+\mathbf{p}\wedge\mathbf{p}
   -(\mathbf{h}\wedge\mathbf{p}+\mathbf{p}\wedge\mathbf{h})-(\mathbf{h}\wedge\mathbf{q}+\mathbf{q}\wedge\mathbf{h})
   +\frac{1}{2}(\mathbf{q}\wedge\mathbf{q}-\mathbf{q}'\wedge\mathbf{q}')\,,\nonumber\\
d\mathbf{p}\,&=&-2\mathbf{p}\wedge\mathbf{p}-(\mathbf{p}\wedge\mathbf{q}+\mathbf{q}\wedge\mathbf{p})
   +\frac{1}{2}(\mathbf{q}\wedge\mathbf{q}+\mathbf{q}'\wedge\mathbf{q}')\,,\nonumber\\[0.2cm]
d\mathbf{q}\,&=&-2\mathbf{q}\wedge\mathbf{q}\,,\nonumber\\[0.2cm]
d\mathbf{q}'&=&-2(\mathbf{p}\wedge\mathbf{q}'+\mathbf{q}'\wedge\mathbf{p})
   -(\mathbf{q}\wedge\mathbf{q}'+\mathbf{q}'\wedge\mathbf{q})\,.
\end{eqnarray}
which are the same as (\ref{MCL}) and also (3.10) of ref. \cite{BPR03}.

The action of the Green-Schwarz superstring in $AdS_3\times S^3$ space is
\begin{eqnarray}
S&=&\int (\,\mathfrak{L}_0+\mathfrak{L}_1)\,,\\
\mathfrak{L}_0&=&-\frac{1}{2}\,(\,L^{a}\wedge\ast L^{a}+L^{a'}\wedge\ast L^{a'})\,,\\[0.1cm]
\mathfrak{L}_1&=&\frac{1}{2}\,(\,\bar{L}^1\wedge
L^2+\bar{L}^2\wedge L^1)\,.
\end{eqnarray}
Here we use a quadratic form of the Wess-Zumino term. The
differential of $\mathfrak{L}_1$ coincides with the WZ term in a
cubic form used in \cite{RR98}. It should be noticed that
$\bar{L}^1\wedge L^2=\bar{L}^2\wedge L^1$ in the $AdS_5\times S^1$
case, while here $\bar{L}^2\wedge L^1$ is the Hermitian
conjugation of $\bar{L}^1\wedge L^2$. It is also not difficult to
check the $\kappa$-symmetry of this action. The variations of the
Maurer-Cartan 1-forms are
\begin{eqnarray}
\delta L^a &=& d\omega^a+L^{ab}\omega^b+L^b\omega^{ba}
           +\frac{\mathrm{i}}{2}(\,\bar{L}^I\gamma^a\omega^I-\bar{\omega}^I\gamma^aL^I)\,,\\
\delta L^{a'}&=&d\omega^{a'}+L^{a'b'}\omega^{b'}
           +L^{b'}\omega^{b'a'}
           -\frac{1}{2}(\,\bar{L}^I\gamma^{a'}\omega^I-\bar{\omega}^I\gamma^{a'}L^I)\,,\\
\delta L^I&=&d\omega^I
           +\frac{1}{2}\varepsilon^{IJ}\omega^{\hat{a}}\,\tilde{\gamma}^{\hat{a}}L^J
           -\frac{1}{2}\varepsilon^{IJ}L^{\hat{a}}\tilde{\gamma}^{\hat{a}}\omega^J
           -\frac{1}{4}L^{\hat{a}\hat{b}}\,\tilde{\gamma}^{\hat{a}\hat{b}}\omega^I
           +\frac{1}{4}\omega^{\hat{a}\hat{b}}\,\tilde{\gamma}^{\hat{a}\hat{b}}L^I,\\[0.1cm]
\delta \bar{L}^I&=&d\bar{\omega}^I
           +\frac{1}{2}\varepsilon^{IJ}\omega^{\hat{a}}\bar{L}^J\tilde{\gamma}^{\hat{a}}
           -\frac{1}{2}\varepsilon^{IJ}L^{\hat{a}}\bar{\omega}^J\tilde{\gamma}^{\hat{a}}
           +\frac{1}{4}L^{\hat{a}\hat{b}}\bar{\omega}^I\tilde{\gamma}^{\hat{a}\hat{b}}
           -\frac{1}{4}\omega^{\hat{a}\hat{b}}\bar{L}^I\tilde{\gamma}^{\hat{a}\hat{b}},
\end{eqnarray}
where $ \tilde{\gamma}^{\hat{a}}=\left\{
\begin{array} {r@{\quad,\quad}l}
   -\mathrm{i}\gamma^a & \hat{a}=a \\
   \gamma^{a'} & \hat{a}=a'
\end{array}\right.\quad \mathrm{and} \quad
\tilde{\gamma}^{\hat{a}\hat{b}}=\left\{
\begin{array} {r@{\quad,\quad}l}
   -\gamma^{ab}& \hat{a}\hat{b}=ab \\
   -\gamma^{a'b'} & \hat{a}\hat{b}=a'b'
\end{array}\right.$\\[0.3cm]
The variation of the Wess-Zumino term is
\begin{eqnarray}
&&\delta\mathcal{L}_1 = \frac{1}{2}\,
      [\,-s^{IJ}\bar{L}^I \omega^{\hat{a}} \tilde{\gamma}^{\hat{a}}\wedge L^J
      +s^{IJ}\bar{\omega}^I L^{\hat{a}} \tilde{\gamma}^{\hat{a}}\wedge L^J\nonumber\\[0.1cm]
&&\quad\quad\quad\quad +s^{IJ}\bar{L}^I \wedge L^{\hat{a}}
\tilde{\gamma}^{\hat{a}}\omega^J
      +\sigma^{IJ}d(\,\bar{\omega}^I L^J-\bar{L}^I \omega^J)\,]\,,
\end{eqnarray}
where $(s^{IJ})=\begin{pmatrix}1&0\\0&-1\end{pmatrix}$
and $(\,\sigma^{IJ})=\begin{pmatrix}0&1\\1&0\end{pmatrix}$.\\[0.2cm]
The equations of motions of $AdS_3\times S^3$ are
\begin{eqnarray}
&&\sqrt{-g}g^{ij}(\nabla_iL^{a}_j+L^{ab}_iL^b_j)
     +\frac{\mathrm{i}}{2}\varepsilon^{ij}s^{IJ}\bar{L}^I_i\gamma^aL^J_j=0,\nonumber\\
&&\sqrt{-g}g^{ij}(\nabla_iL^{a'}_j+L^{a'b'}_iL^{b'}_j)
     -\frac{1}{2}\varepsilon^{ij}s^{IJ}\bar{L}^I_i\gamma^{a'}L^J_j=0,\nonumber\\[0.1cm]
&&(L^a_i\gamma^a+\mathrm{i}L^{a'}_i\gamma^{a'})
     (\sqrt{-g}g^{ij}\delta^{IJ}-\varepsilon^{ij}s^{IJ})L^J_j=0,\nonumber\\[0.2cm]
&&\bar{L}^I_i(L^a_i\gamma^a+\mathrm{i}L^{a'}_i\gamma^{a'})
     (\sqrt{-g}g^{ij}\delta^{IJ}+\varepsilon^{ij}s^{IJ})=0.
\end{eqnarray}
In terms of the lowercase 1-forms the equations of motions are
translated into
\begin{eqnarray}
d\ast\mathbf{p}&=&(\,\mathbf{p}\wedge\ast\,\mathbf{q}+\ast\,\mathbf{q}\wedge\mathbf{p})
  +\frac{1}{2}(\mathbf{q}\wedge\mathbf{q}'+\mathbf{q}'\wedge\mathbf{q})\,,\nonumber\\[0.05cm]
0&=&\mathbf{p}\wedge(\ast\,\mathbf{q}-\mathbf{q}')+(\ast\,\mathbf{q}-\mathbf{q}')\wedge\mathbf{p}\,,\nonumber\\[0.1cm]
0&=&\mathbf{p}\wedge(\mathbf{q}-\ast\,\mathbf{q}')+(\mathbf{q}-\ast\,\mathbf{q}')\wedge\mathbf{p}\,.
\end{eqnarray}
which also have the same form as (\ref{EQL}) and the ones of the
$AdS_5\times S^5$ string. Because of the $\mathbb{Z}_4$-grading
structure and the similar forms of the equations of motions the
existence of a family of one-parameter flat currents is evident,
and their expressions in terms of the lowercase 1-forms are the
same as the ones of $AdS_5\times S^5$ string.\footnote{The
nonlocal conserved currents in $AdS_3\times S^3$ \emph{NSR}-superstring
has been studied in reference \cite{Maha05}.}


\section{Discussions}

In this paper we construct a one-parameter family of  nonlocal
currents from the $\kappa$-symmetric actions of Green-Schwarz
superstring in both $AdS_5\times S^1$ and $AdS_3\times S^3$
backgrounds. In the former case, although it is briefly described
in \cite{Poly04}, here we use a new $\mathbb{Z}_4$-graded base of
the superalgera $su(2,2|2)$ to rewrite the action. We take the
quadratic Wess-Zumino term and determine the coefficient by
requiring the $\kappa$-symmetry of the whole action. Then we
construct explicitly a one-parameter family of flat currents
which would lead to classical nonlocal charges. This fact suggests
that the sigma model of $AdS_5\times S^1$ string is really
completely integrable, just as being asserted in \cite{Poly04}.
Similarly, we discuss the construction of nonlocal currents in
$AdS_3 \times S^3$ superstring, whose symmetry algebra also has a
$\mathbb{Z}_4$ grading. From the construction, we notice that in
terms of the $\mathcal{G}_0$-gauge invariant lowercase forms, the
Maurer-Cartan equations and the equations of motion take the same
form as the ones in $AdS_5\times S^5$ case. As a  result,   the
nonlocal flat currents in all the three supercoset models with
$\mathbb{Z}_4$ grading take the similar form.

It is a notorious fact that the superstring in $AdS_5\times S^5$
with RR-backgrounds is difficult to be quantized. One promising
approach is the covariant quantization in the pure spinor
formalism\cite{Berk03}.  It would be very interesting to
investigate whether one can have a pure spinor formulation for
$AdS_5\times S^1$ superstring and check whether the nonlocal
charges survive the quantization.

On the other hand, it would be illuminating to set up the
dictionary in the correspondence between the noncritical
superstring in $AdS_5\times S^1$ space and $\mathcal{N}=1$
super-Yang-Mills with flavors. Actually, the pure $\mathcal{N}=1$
super-Yang-Mills could not be conformal invariant and so cannot be
the dual of closed superstring in $AdS_5\times S^1$ spacetime.
However, the recent discovery in \cite{KM04} implies that a
superstring theory with closed and open string degrees of freedom
in $AdS_5\times S^1$ spacetime, which is different from the one
studied in this paper, may dual to a conformal invariant
$\mathcal{N}=1$ super-Yang-Mills with flavors.  The correspondence
and the possible role played by the integrable structure deserves
further investigation.


\section*{Note}

While we are finishing the paper, there appears a
paper\cite{Young} in arXiv, which has some overlap with our paper.
Our result is consistent with its argument. One remarkable fact is
that in \cite{Young}, the coefficient of Wess-Zumino term is
determined by requiring the existence of a one-parameter family
of flat currents, without taking into account of $\kappa$-symmetry.
In this paper, we treat the $\kappa$-symmetry seriously and fix
Wess-Zumino term from it. It seems that the $\kappa$-symmetry
implies the existence of the flat currents. The relation between
$\kappa$-symmetry and flat currents deserves further study.

\section*{Acknowledgements}
The work of PZ was supported by a grant of NSFC and a grant of
Chinese Academy of Sciences, the work of BC was supported by NSFC
grant 10405028 and a grant of Chinese Academy of Sciences.

\appendix
\section{Notation}
The convention of this paper is listed in this appendix.
The barred Latin letters $\bar{a},\bar{b}=0,1,2,3,4,5$ are the $so(4,2)$ vector indices.
The ordinary Latin letters $a,b=0,1,2,3,4$ are the $so(4,1)$ vector indices ($AdS_5$ tangent space).
The primed Latin letters $a',b'=1,2,3$ are the $so(3)$ vector indices.
The ordinary Greek letters $\alpha,\beta=1,2,3,4$ are the $so(4,2)$ spinor indices
(also the $so(4,1)$ spinor indices).
The primed Greek letters $\alpha',\beta'=1,2$ are the $so(3)$ spinor indices.
The six-dimensional metric is $(\eta_{\bar{a}\bar{b}})=\mathrm{diag}(-++++\,\,-)$,
the five-dimensional metric is $(\eta_{ab})=\mathrm{diag}(-++++)$,
and the four-dimensional metric is $(\eta_{\mu\nu})=\mathrm{diag}(-++\,+)$.

We choose the representation of 4-dimensional Dirac matrices $\gamma^\mu$ as following
\begin{eqnarray}
\gamma^\mu=-\mathrm{i}\begin{pmatrix}0&\sigma^\mu\\-\bar{\sigma}^\mu& 0\end{pmatrix}\quad\mathrm{with}\quad
\sigma^\mu=(\mathbf{1},\sigma^i)\quad \mathrm{and} \quad \bar{\sigma}^\mu=(\mathbf{1},-\sigma^i).
\end{eqnarray}
They satisfy the standard anticommutation relations
$
\{\gamma^\mu,\gamma^\nu\}=2\eta^{\mu\nu},
$
and $\sigma^i$ are three Pauli matrices. Then we define the $\bar{\gamma}$ as
\begin{eqnarray}
\bar{\gamma}=-\gamma^0\gamma^1\gamma^2\gamma^3=\begin{pmatrix}-1 & 0\\0 & 1\end{pmatrix}.
\end{eqnarray}
Now we use the above $\gamma_\mu$ and $\bar{\gamma}$ to construct the generators
$\Gamma_{\bar{a}\bar{b}}$ of $so(4,2)$:
\begin{eqnarray}
&\Gamma_{\mu\nu}=\frac{1}{2}[\gamma_\mu,\gamma_\nu]  -\frac{1}{2}\begin{pmatrix}\sigma_{\mu\nu}&0\\0&\bar{\sigma}_{\mu\nu}\end{pmatrix},&\quad
  \Gamma_{\mu 4}=\gamma_\mu \bar{\gamma}  -\mathrm{i}\begin{pmatrix}0&\sigma_\mu\\-\bar{\sigma}_\mu& 0\end{pmatrix},\nonumber\\[0.2cm]
&\Gamma_{\mu 5}=\gamma_\mu  -\mathrm{i}\begin{pmatrix}0&\sigma_\mu\\\bar{\sigma}_\mu&0\end{pmatrix},&\quad
  \Gamma_{45}=\bar{\gamma}=\begin{pmatrix}-1 & 0\\0 & 1\end{pmatrix}.
\end{eqnarray}
It is not difficult to check that $\Gamma_{\bar{a}\bar{b}}$ satisfies the standard commutation relations:
\begin{eqnarray}
\left[\,\frac{1}{2}\,\Gamma_{\bar{a}\bar{b}}\,,\,\frac{1}{2}\,\Gamma_{\bar{c}\bar{d}}\,\right]
=\frac{1}{2}\eta_{\bar{a}\bar{d}}\,\Gamma_{\bar{b}\bar{c}}+\frac{1}{2}\eta_{\bar{b}\bar{c}}\,\Gamma_{\bar{a}\bar{d}}
-\frac{1}{2}\eta_{\bar{a}\bar{c}}\,\Gamma_{\bar{b}\bar{d}}-\frac{1}{2}\eta_{\bar{b}\bar{d}}\,\Gamma_{\bar{a}\bar{c}}\,.
\end{eqnarray}

When we construct the graded generators of $su(2,2|2)$ we split the $so(4,1)$ 4-spinors
into its 2-component forms:
\begin{eqnarray}
\Psi_\alpha=(\phi^A\,,\,\chi_{\dot{A}}\,).
\end{eqnarray}
Here $A,B=1,2$ and $\dot{A},\dot{B}=3,4$. Define
\begin{eqnarray}
&&(\varepsilon_{AB})=(\varepsilon_{\dot{A}\dot{B}})=\begin{pmatrix}0&1\\-1&0\end{pmatrix},\quad
  (\varepsilon^{AB})=(\varepsilon^{\dot{A}\dot{B}})=\begin{pmatrix}0&-1\\1&0\end{pmatrix}.
\end{eqnarray}
So we have
\begin{eqnarray}
\varepsilon^{AB}\varepsilon_{BC}=\delta^A_C\,,\quad
\varepsilon^{\dot{A}\dot{B}}\varepsilon_{\dot{B}\dot{C}}=\delta^{\dot{A}}_{\dot{C}}\,.
\end{eqnarray}
The 2-spinor indices can be raised and lowered by the above $\varepsilon$-tensors
\begin{eqnarray}
&&\psi^A=\varepsilon^{AB}\psi_B,\quad
  \psi_A=\varepsilon_{AB}\psi^B;\nonumber\\
&&\psi^{\dot{A}}=\varepsilon^{\dot{A}\dot{B}}\psi_{\dot{B}},\quad
  \psi_{\dot{A}}=\varepsilon_{\dot{A}\dot{B}}\psi^{\dot{B}}.
\end{eqnarray}
The index structures of $\sigma^\mu$ and $\bar{\sigma}^\mu$ are
\begin{eqnarray}
\sigma^\mu=(\sigma^\mu)^{A\dot{B}},\quad \bar{\sigma}^\mu=(\bar{\sigma}^\mu)_{\dot{A}B}.
\end{eqnarray}
It is true that
\begin{eqnarray}
(\bar{\sigma}^\mu)_{\dot{A}B}=(\sigma^\mu)_{B\dot{A}}
\equiv\varepsilon_{BC}\,\varepsilon_{\dot{A}\dot{D}}\,(\sigma^\mu)^{C\dot{D}} .
\end{eqnarray}
We can use the $\varepsilon$-tensors to construct the $4\times4$ charge conjugation matrix of $so(4,1)$
Clifford algebra
\begin{eqnarray}
C=(C_{\alpha\beta})=\begin{pmatrix}\varepsilon_{AB}&0\\0&-\varepsilon^{\dot{A}\dot{B}}\end{pmatrix},\label{C}
\end{eqnarray}
which has the characteristic property that
\begin{eqnarray}
C\gamma^aC^{-1}=(\gamma^a)^T.
\end{eqnarray}
We also define the $2\times2$ charge conjugation matrix of $so(3)$ Clifford algebra
\begin{eqnarray}
C\,'=(C_{\alpha'\beta'})=\begin{pmatrix}0&1\\-1&0\end{pmatrix},\label{C'}
\end{eqnarray}
which has the similar property
\begin{eqnarray}
C\,'\tau^{a'}(C\,')^{-1}=-(\tau^{a'})^T.
\end{eqnarray}

\section{$\kappa$-symmetry of the $AdS_5\times S^1$ superstring}
Now we check that when $k=-2$ the whole action $S=S_0 + S_1$
is invariant under the local $\kappa$-transformations
\begin{eqnarray}
&&\omega^a_{\kappa}=\omega^R_{\kappa}=0\,,\\[0.1cm]
&&\omega^1_{\kappa}=P_-^{ij}(L^a_i\gamma^a-\mathrm{i} L^R_i)\kappa^1_j\,,
  \quad\omega^2_{\kappa}=P_+^{ij}(L^a_i\gamma^a-\mathrm{i} L^R_i)\kappa^2_j\,,\\[0.2cm]
&&\delta_\kappa(\sqrt{-g}g^{ij})
  =-8\mathrm{i}\sqrt{-g}\,(P_-^{il}P_-^{jk}\bar{L}^1_k\kappa^1_l
  +P_+^{il}P_+^{jk}\bar{L}^2_k\kappa^2_l).
\end{eqnarray}
Here we have defined two projectors
$P_\pm^{ij}=\frac{1}{2}(g^{ij}\pm\frac{1}{\sqrt{-g}}\varepsilon^{ij})$.
They have the following useful properties
\begin{eqnarray}
&& P_\pm^{ij}=P_\mp^{ji}\,,\label{Psym}\\
&& P_\pm^{ij}P_\pm^{kl}=P_\pm^{kj}P_\pm^{il}\,. \label{PP}
\end{eqnarray}
Let $k=-2$ in the action. By using the variations of the Maurer-Cartan 1-forms we have
\begin{eqnarray}
\delta_\kappa S&=&\int d^2\sigma\,(\Delta_1+\Delta_2)\,,\\
\Delta_1&=&-\frac{1}{2}\delta_\kappa(\sqrt{-g}g^{ij})\,(L^a_iL^a_j+L^R_iL^R_j)\,,\\[0.2cm]
\Delta_2&=&-2\mathrm{i}\,(\sqrt{-g}g^{ij}\delta^{IJ}+\varepsilon^{ij}s^{IJ})\,
          \bar{L}^I_i(L^a_j\gamma^a+\mathrm{i} L^R_j)\,\omega^J_{\kappa}\,, \nonumber\\[0.2cm]
&=&-4\mathrm{i}\sqrt{-g}\left[\,P_+^{ij}P_-^{kl}\bar{L}^1_i(L^a_j\gamma^a+\mathrm{i} L^R_j)(L^b_k\gamma^b-\mathrm{i} L^R_k)\kappa^1_l\right.\nonumber\\[0.1cm]
& &\qquad\qquad\quad\left.
   +\,P_-^{ij}P_+^{kl}\bar{L}^2_i(L^a_j\gamma^a+\mathrm{i} L^R_j)(L^b_k\gamma^b-\mathrm{i} L^R_k)\kappa^2_l\,\right]\,.
\end{eqnarray}
By virtue of (\ref{Psym}) (\ref{PP}) it is not very difficult to check that
$\Delta_2$ is just the opposite of $\Delta_1$. So the whole action has the $\kappa$-
symmetry when the coefficient $k$ in front of the Wess-Zumino term is set to be $-2$.

\section{Maurer-Cartan 1-forms of $SU(2,2|2)/(SO(4,1)\times SO(3))$}
We can use the methods of \cite{KRR98} to write down the explicit forms of
$L^a,L^R,L^I$ as follows
\begin{eqnarray}
L^a&=&e^a_\mu dx^\mu-\mathrm{i}\,\bar{\theta}^I\gamma^a
    \left[\left(\frac{\sinh\mathcal{M}/2}{\mathcal{M}/2}\right)^2 D\theta\right]^I\,,\\[0.1cm]
L^R&=&e^R_\mu dx^\mu-\mathrm{i}\,\bar{\theta}^I
    \left[\left(\frac{\sinh\mathcal{M}/2}{\mathcal{M}/2}\right)^2 D\theta\right]^I\,,\\[0.2cm]
L^I&=&\left[\left(\frac{\sinh\mathcal{M}}{\mathcal{M}}\right)D\theta\right]^I\,,
\end{eqnarray}
where
\begin{eqnarray}
(\mathcal{M}^2)^{IJ}=\varepsilon^{IK}(-\gamma^a\theta^K\bar{\theta}^J\gamma^a+\theta^K\bar{\theta}^J)
   +\frac{1}{2}\,\varepsilon^{KJ}(\gamma^{ab}\theta^I\bar{\theta}^K\gamma^{ab}-4\tau^{a'}\theta^I\bar{\theta}^K\tau^{a'})\,,\label{M}
\end{eqnarray}
and
\begin{eqnarray}
(D\theta)^I=d\theta^I+\frac{1}{4}(A^{ab}\gamma^{ab}+2A^{a'}\tau^{a'})\,\theta^I
   -\frac{\mathrm{i}}{2}\,\varepsilon^{IJ}(e^a\gamma^a+\mathrm{i} e^R)\,\theta^J\,.
\end{eqnarray}
Here $e^a$ and $e^R$ are the bosonic vielbeins corresponding to
$AdS_5$ directions and $S^1$ directions, $A^{ab}$ and $A^{a'}$ are
the bosonic connections.
Notice that the last term in (\ref{M}) is different from the $AdS_5\times S^5$
case, in which the coefficient is $-1$, not $-4$.
Using these formulae we can find the $\theta$-expansion of the
action of $AdS_5\times S^1$ superstring.

\end{document}